\documentclass[preprints,article,accept,moreauthors,pdftex,10pt,a4paper]{Definitions/mdpi} 

\firstpage{1} 
\pubvolume{xx}
\issuenum{1}
\articlenumber{1}
\pubyear{2018}
\copyrightyear{2018}
\history{}

\usepackage[english]{babel}

\usepackage{booktabs}

\usepackage[ruled,linesnumbered]{algorithm2e}
\usepackage{subcaption}
\usepackage{xcolor}

\usetikzlibrary{decorations.pathreplacing,patterns}
\newlength{\hatchspread}
\newlength{\hatchthickness}
\newlength{\hatchshift}
\newcommand{\hatchcolor}{}
\tikzset{hatchspread/.code={\setlength{\hatchspread}{#1}},
	hatchthickness/.code={\setlength{\hatchthickness}{#1}},
	hatchshift/.code={\setlength{\hatchshift}{#1}},
	hatchcolor/.code={\renewcommand{\hatchcolor}{#1}}}
\tikzset{hatchspread=3pt,
	hatchthickness=0.4pt,
	hatchshift=0pt,
	hatchcolor=black}
\pgfdeclarepatternformonly[\hatchspread,\hatchthickness,\hatchshift,\hatchcolor]
{custom north west lines}
{\pgfqpoint{\dimexpr-2\hatchthickness}{\dimexpr-2\hatchthickness}}
{\pgfqpoint{\dimexpr\hatchspread+2\hatchthickness}{\dimexpr\hatchspread+2\hatchthickness}}
{\pgfqpoint{\dimexpr\hatchspread}{\dimexpr\hatchspread}}
{
	\pgfsetlinewidth{\hatchthickness}
	\pgfpathmoveto{\pgfqpoint{0pt}{\dimexpr\hatchspread+\hatchshift}}
	\pgfpathlineto{\pgfqpoint{\dimexpr\hatchspread+0.15pt+\hatchshift}{-0.15pt}}
	\ifdim \hatchshift > 0pt
	\pgfpathmoveto{\pgfqpoint{0pt}{\hatchshift}}
	\pgfpathlineto{\pgfqpoint{\dimexpr0.15pt+\hatchshift}{-0.15pt}}
	\fi
	\pgfsetstrokecolor{\hatchcolor}
	\pgfusepath{stroke}
}

\Title{Dynamic Load Balancing Techniques for Particulate Flow Simulations}

\Author{Christoph Rettinger $^{1,}$*\orcidA{} and Ulrich Rüde $^{1,2}$\orcidB{}}

\AuthorNames{Christoph Rettinger and Ulrich Rüde}

\address{%
$^{1}$ \quad Chair for System Simulation, Friedrich--Alexander--Universität Erlangen--Nürnberg, Cauerstraße 11, 91058 Erlangen, Germany\\
$^{2}$ \quad CERFACS, 42 Avenue Gaspard Coriolis, 31057 Toulouse Cedex 1, France}

\corres{Correspondence: christoph.rettinger@fau.de}

\abstract{
	Parallel multiphysics simulations often suffer from load imbalances originating from the applied coupling of algorithms with spatially and temporally varying workloads.
	It is thus desirable to minimize these imbalances to reduce the time to solution and to better utilize the available hardware resources.
	Taking particulate flows as an illustrating example application, we present and evaluate load balancing techniques that tackle this challenging task.
	This involves a load estimation step in which the currently generated workload is predicted.
	We describe in detail how such a workload estimator can be developed.
	In a second step, load distribution strategies like space-filling curves or graph partitioning are applied to dynamically distribute the load among the available processes.
	To compare and analyze their performance, we employ these techniques to a benchmark scenario and observe a reduction of the load imbalances by almost a factor of four.
	This results in a decrease of the overall runtime by 14\% for space-filling curves.
}


\begin{document}

\section{Introduction}

Simulations are becoming an increasingly prominent alternative to often expensive and time consuming laboratory experiments
Therefore, engineers are interested in introducing more and more physical effects to explore complex phenomena with the help of computers.
These multiphysics simulations thus feature a combination of different algorithms for the physics, like multiphase fluid flow \cite{kusumaatmaja_hemingway_fielding_2016}, particle motion \cite{Rettinger2017ISC}, and free surfaces \cite{ANDERL2014331}.
Due to their computational cost, it is often necessary to utilize thread-based and distributed memory parallelization techniques to obtain results in reasonable time and to make use of the immense capabilities provided by today's supercomputers.
This is commonly achieved by a spatial domain partitioning where the computational domain is subdivided into smaller pieces which are then distributed among the available processes, see e.g. \cite{Godenschwager2013}.
As the majority of the parallel numerical codes consist of computation and synchronization, the desired properties of such a distribution are that the overall workload (originating from the computations) is spread evenly among the processes and that the amount of communication (due to synchronization) is minimized.
If the first property is violated, the simulation suffers from load imbalances and the processes must wait at the synchronization points for the slowest process, i.e. the one with the largest workload, until the simulation can proceed.
As a result, the runtime of the whole simulation increases and the hardware resources at hand are utilized less efficiently due to the idle times.
Acquiring and maintaining a distribution with balanced workloads is thus crucial to achieve efficient parallel simulations.

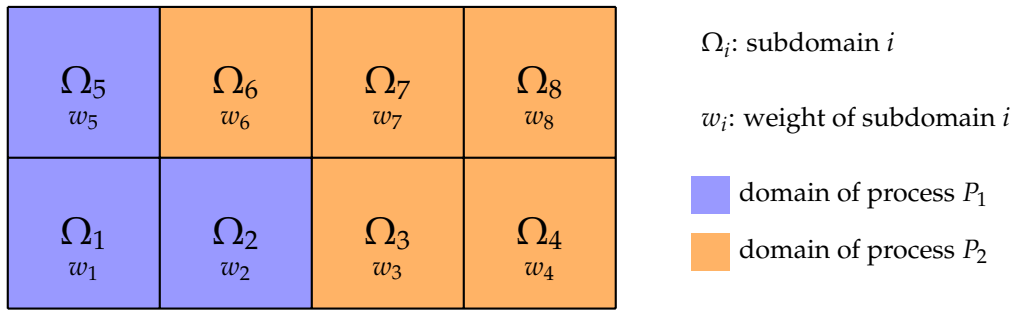
\begin{figure}[tb]
	\centering
	\begin{tikzpicture}
	\fill[blue!40!white] (0,0) rectangle ++(2,2);
	\fill[blue!40!white] (2,0) rectangle ++(2,2);
	\fill[blue!40!white] (0,2) rectangle ++(2,2);
	\fill[orange!60!white] (4,0) rectangle ++(2,2);
	\fill[orange!60!white] (6,0) rectangle ++(2,2);
	\fill[orange!60!white] (2,2) rectangle ++(2,2);
	\fill[orange!60!white] (4,2) rectangle ++(2,2);
	\fill[orange!60!white] (6,2) rectangle ++(2,2);
	\draw[step=2,black,thick] (0,0) grid (8,4);
	\node at (1,1) {\Large$\Omega_1$};
	\node at (3,1) {\Large$\Omega_2$};
	\node at (5,1) {\Large$\Omega_3$};
	\node at (7,1) {\Large$\Omega_4$};
	\node at (1,3) {\Large$\Omega_5$};
	\node at (3,3) {\Large$\Omega_6$};
	\node at (5,3) {\Large$\Omega_7$};
	\node at (7,3) {\Large$\Omega_8$};
	\node at (1,0.5) {$w_1$};
	\node at (3,0.5) {$w_2$};
	\node at (5,0.5) {$w_3$};
	\node at (7,0.5) {$w_4$};
	\node at (1,2.5) {$w_5$};
	\node at (3,2.5) {$w_6$};
	\node at (5,2.5) {$w_7$};
	\node at (7,2.5) {$w_8$};
	\node[right] at (9,3.5) {$\Omega_i$: subdomain $i$};
	\node[right] at (9,2.5) {$w_i$: weight of subdomain $i$};
	\fill[blue!40!white] (9,1.25) rectangle ++(0.5,0.5);
	\node[right] at (9.5,1.5) {domain of process $P_1$};
	\fill[orange!60!white] (9,0.5) rectangle ++(0.5,0.5);
	\node[right] at (9.5,0.75) {domain of process $P_2$};
	\end{tikzpicture}
	\caption{Illustration of the domain partitioning with load balancing. The computational domain $\Omega$ is subdivided into subdomains $\Omega_i$, such that $\Omega=\bigcup_i\Omega_i$. The load estimator assigns a weight $w_i$ to each subdomain that quantifies the workload. The load distribution routine then assigns the subdomains to the available processes (here $P_1$ and $P_2$) such that the load per process is balanced.}
	\label{fig:domain_part_blocks}
\end{figure}

Conceptually, the task of load balancing can be split into two steps:
At first, referred to as \textit{load estimation}, each subdomain is assigned a weight that quantifies the workload present on that subdomain.
Based on these weights, a \textit{load distribution} routine is executed in a second step to reassign the subdomains to the processes.
An illustration is given in Figure~\ref{fig:domain_part_blocks}.
Depending on the characteristics of the underlying algorithms and the targeted degree of parallelism, various load balancing strategies have been developed, as reviewed in \cite{HENDRICKSON2000485}.
Meshless methods, like particle simulations, typically use the number of particles \cite{owen2000,Kloss2012} or collisions \cite{eibl2018systematic} per subdomain as weight function.
Consequently, this weight may vary significantly in space and time, requiring dynamic load balancing techniques.
This can be achieved, e.g., by changing the geometric position of the subdomains \cite{owen2000,Kloss2012} or further subdivision \cite{eibl2018systematic}. 
Contrary to that, in grid-based algorithms like finite volume or lattice Boltzmann methods, the weight of a subdomain is typically given as the number of grid cells \cite{deiterding2011block,LINTERMANN2014131,Schornbaum2016,Qi2016}.
As a special case, if the grid remains unchanged during the simulation and a constant number of cells per subdomain is chosen, a distribution of one subdomain per process is an obvious choice and thus encountered in many such programs.
However, when adaptive mesh refinement (AMR) is applied to improve the accuracy and efficiency of these simulations \cite{Deiterding2016,Schornbaum2018}, the workload is altered significantly and load balancing becomes an essential component.
Apart from these techniques, load distribution routines have attracted a lot of attention throughout the years and various methods have been proposed, since they are also used for classical graph partitioning problems \cite{HENDRICKSON2000485, bader2012space, Buluc2016}.

It becomes apparent, however, that the applied techniques differ substantially for meshless and grid-based methods where in the first case the subdomains are modified solely for load balancing whereas this is not desired for the grid-based methods.
Determining a suitable load balancing approach for multiphysics simulations that incorporate both types of algorithms thus poses a challenge.
Commonly, in cases where one of the algorithms is the workload-wise dominant one, the specific approaches of this algorithm are used, e.g. the grid-based fluid solver in particulate flow simulations \cite{schneiders2015efficient,Rettinger2017ISC}, which comes at the cost of load imbalances.
Alternatively, a different domain partitioning for each part can be used \cite{owen2000}, which, however, requires expensive communication and complicated mapping mechanisms between the subdomains for distributed memory simulations.
In this paper, we will therefore present a different strategy for dynamic load balancing of these simulations.
We will use particulate flows as an illustrating example, which features a coupling between a fluid solver and a particle simulation.
Our primary design objective is to enable massively parallel simulations on supercomputers for large physical systems.
This restriction already excludes all algorithms that require global knowledge about process local quantities, e.g. the position of all particles, as they will inevitably not scale to several thousand or more processes.
Additionally, we require that all parts of our simulation use the same domain partitioning to avoid the previously mentioned difficulties and bottlenecks.
A variant that complies with these specifications is a static partitioning of the computational domain into blocks of constant size \cite{Godenschwager2013}, as shown in Figure~\ref{fig:domain_part_blocks}.
Load balancing is achieved by dynamically distributing these blocks among the available processes with the goal to have a similar workload on each process by specifically assigning several blocks to each process.
We develop a genuine load estimation approach to quantify the workload per block by taking into account all aspects of the coupled simulation and evaluate different load distribution routines in detail.

The remainder of this paper is thus structured as follows: 
At first, we briefly describe the numerical methods we apply for geometrically fully resolved particulate flow simulations in Sec.~\ref{sec:numerical_method}, consisting of the lattice Boltzmann method, a hard contact particle solver and the fluid-particle coupling mechanism.
Next, in Sec.~\ref{sec:load_determination}, we present and calibrate our load estimation strategy that predicts the block's weight based on locally available quantities.
This estimator is then applied in Sec.~\ref{sec:comparison_load_distribution}, where additionally the performance of several load distribution approaches is investigated and compared.
We summarize our findings and give an outlook on future directions in Sec.~\ref{sec:conclusion}.

\section{Numerical Methods}
\label{sec:numerical_method}

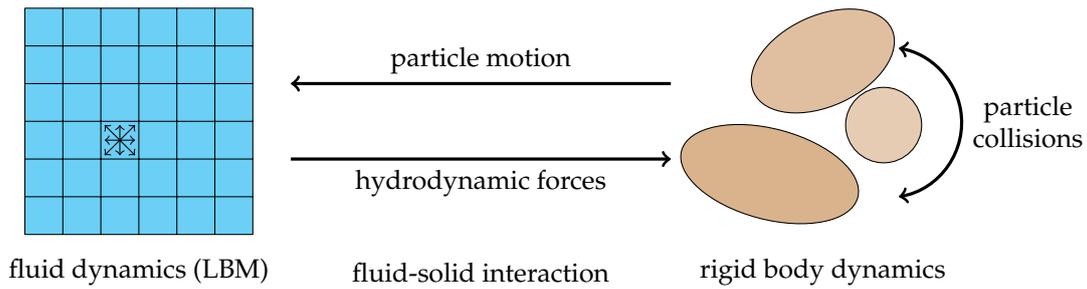
\begin{figure}[tb]
\centering
\begin{tikzpicture}
\fill[cyan!50!white] (-1.5,0) rectangle ++(3,3);
\draw[step=0.5,black,thin] (-1.5,0) grid ++(3,3);
\draw[<->] (-0.45,1.05) --++ (0.4,0.4);
\draw[<->] (-0.45,1.25) --++ (0.4,0);
\draw[<->] (-0.25,1.05) --++ (0,0.4);
\draw[<->] (-0.05,1.05) --++ (-0.4,0.4);
\node at (0,-0.5) {fluid dynamics (LBM)};

\draw[fill=brown!50!white,shift={(9,2.3)},rotate=25] (0,0) ellipse (1 and 0.6);
\draw[fill=brown!40!white] (9.8,1.45) circle (0.5);
\draw[fill=brown!60!white,shift={(8.3,0.8)},rotate=-15] (0,0) ellipse (1.2 and 0.6);
\draw[<->,very thick] (10,0.5) arc (-80:80:1);
\node[align=center] at (11.7,1.5) {particle\\ collisions};
\node at (9,-0.5) {rigid body dynamics};

\draw[->,very thick] (2,1) --++ (5,0) node[pos=0.5,anchor=north] {hydrodynamic forces};
\draw[->,very thick] (7,2) --++ (-5,0) node[pos=0.5,anchor=south] {particle motion};
\node at (4.5,-0.5) {fluid-solid interaction};
\end{tikzpicture}
\caption{Schematic representation of the numerical methods presented in Sec.~\ref{sec:numerical_method} that are used for particulate flow simulations in this work.}
\label{fig:numerical_methods}
\end{figure}

In this section, we briefly describe the numerical methods that we apply in our simulations of particulate flows.
The fluid flow is computed via the lattice Boltzmann method and a non-smooth granular dynamics simulation accounts for the motion and collisions of the suspended particles.
A coupling of these two approaches is established based on the momentum exchange principle which features geometrically fully resolved particle shapes \cite{Rettinger2017ISC,RETTINGER201774}.
This coupling is illustrated in Figure~\ref{fig:numerical_methods}.
We note, however, that most of the observations and results obtained in the following sections also carry over to other CFD methods, like finite volume methods, other particle collision resolution methods, like the discrete element method, and also other coupling algorithms, like the immersed boundary method.  

\subsection{Lattice Boltzmann Method}
\label{sec:LBM}

In the lattice Boltzmann method (LBM) \cite{chen_lattice_1998,kruger2017lattice}, the three-dimensional domain is discretized with a uniform lattice.
Each cell features $q$ different particle distribution functions (PDF) $f_q$ where each one is associated with a lattice velocity $\boldsymbol{c}_q$.
We employ the \textit{D3Q19} variant such that $q\in\{0...18\}$.
The macroscopic quantities, density $\rho_f = \rho_0 + \delta \rho_f$, with a mean density $\rho_0$ and a fluctuation $\delta \rho_f$, and fluid velocity $\boldsymbol{u}_f$, are obtained as moments of the PDFs in each grid cell $\boldsymbol{x}$:
\begin{equation}
\delta \rho_f (\boldsymbol{x}, t) = \sum_q f_q(\boldsymbol{x}, t), \ \boldsymbol{u}_f(\boldsymbol{x}, t) = \tfrac{1}{\rho_0} \sum_q \boldsymbol{c}_q f_q(\boldsymbol{x}, t) \label{eq:LBM_Moments}
\end{equation}

The evolution of the PDFs is described by alternating between a collision and a stream step.
We use the two-relaxation-time (TRT) collision operator \cite{ginzburg_two-relaxation-time_2008} that results in the collision step
\begin{equation}
\tilde{f}_q(\boldsymbol{x},t) = f_q(\boldsymbol{x},t)  -\tfrac{1}{\tau_+}\big(f_q^+(\boldsymbol{x},t) - f_q^{\text{eq},+}(\rho_f, \boldsymbol{u}_f)\big) - \tfrac{1}{\tau_-}\big(f_q^-(\boldsymbol{x},t) - f_q^{\text{eq},-}(\rho_f, \boldsymbol{u}_f)\big), \label{eq:LBM_Collide}
\end{equation}
where the PDFs are split into their symmetric $f_q^+$ and anti-symmetric $f_q^-$ parts with their respective collision times $\tau_+$ and $\tau_-$.
This collision step relaxes the PDFs linearly towards their equilibrium values
\begin{equation}
f_q^{\text{eq}}(\rho_f,\boldsymbol{u}) = w_q \left( \rho_f + \rho_0 \left(3\boldsymbol{c}_q \cdot \boldsymbol{u}_f + \tfrac{9}{2}(\boldsymbol{c}_q \cdot \boldsymbol{u}_f)^2 - \tfrac{3}{2}\boldsymbol{u}_f \cdot \boldsymbol{u}_f\right) \right),  \label{eq:LBM_EQ}
\end{equation}
where the lattice weights $w_q$ are taken from \cite{qian_lattice_1992}. 
The succeeding streaming step is then given as
\begin{equation}
f_q( \boldsymbol{x} + \boldsymbol{c}_q \Delta t, t + \Delta t) = \tilde{f}_q(\boldsymbol{x},t). \label{eq:LBM_Stream}
\end{equation}

The kinematic fluid viscosity $\nu$ is determined by the collision time and we use the following relations to define $\tau_+$ and $\tau_-$ \cite{ginzburg_two-relaxation-time_2008}:
\begin{equation}
\nu = \tfrac{1}{3}\left(\tau_+ - \tfrac{\Delta t}{2}\right), \ \left(\tfrac{1}{2} - \tfrac{\tau_+}{\Delta t}\right)\left(\tfrac{1}{2} - \tfrac{\tau_-}{\Delta t}\right) = \tfrac{3}{16}.
\end{equation}

\subsection{Rigid Body Dynamics}
\label{sec:NSGD}
The particles suspended in the fluid interact with each other and walls via collisions.
In order to account for these particle-particle or particle-wall collisions, we apply a rigid body solver which determines the collision forces and integrates the particles' position and velocity in time.
The individual particles are thus represented by their exact geometric shape.
Each particle is described by its position $\boldsymbol{X}_p$, orientation $\boldsymbol{Q}_p$, as well as its translational and rotational velocity, $\boldsymbol{V}_p$ and $\boldsymbol{W}_p$.
The dynamics are described by the Newton-Euler equations for rigid bodies, given acting forces $\boldsymbol{F}_p$ and torques $\boldsymbol{T}_p$.
These arise from external forces like gravity, hydrodynamic interactions with the surrounding fluid, and collisions. 
Here, the collisions are determined by a contact detection algorithm and are modeled as inelastic hard contacts \cite{Preclik2015}, resulting in non-smooth functions for position and orientation.
This effectively resolves overlaps between the particles.
Additionally, the contact model includes Coulomb friction.
The integration of the particle trajectories in time is carried out by a semi-implicit Euler method.
As a result, a non-linear system of equations has to be solved for each particle and each contact \cite{Preclik2015}.
More details about the algorithm and its implementation can be found in \cite{Preclik2014,Preclik2015}.

\subsection{Fluid-solid Interaction}
\label{sec:Coupling}

In order to realize a hydrodynamic coupling between the fluid and the geometrically fully resolved particulate phase, we make use of the momentum exchange method \cite{aidun_direct_1998}.
Here, the particles are explicitly mapped into the fluid domain, marking cells contained inside the particle as \textit{solid}, in contrast to the \textit{fluid} cells.

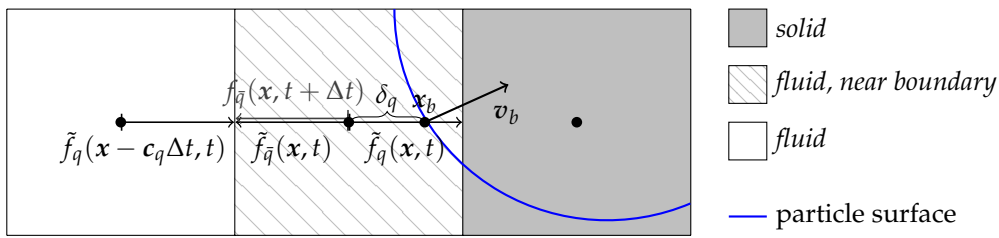
\begin{figure}[htb]
	\centering
	\begin{tikzpicture}
	\fill[lightgray] (6.0,0.0) rectangle (9.0,3.0);
	\fill[pattern = custom north west lines, hatchspread=10pt, hatchcolor=lightgray] (3.0,0.0) rectangle (6.0,3.0);
	\draw[step=3,black,thin] (0,0) grid (9.0,3.0);
	\draw[blue,thick] (5.1,3.0) arc (180:293:2.8);
	\draw[|->,semithick] (4.5,1.5) -- (5.99,1.5) node[pos=0.5,anchor=north] {$\tilde{f}_q(\boldsymbol{x},t)$};
	\draw[|->,semithick] (4.5,1.5) -- (3.01,1.5) node[pos=0.5,anchor=north] {$\tilde{f}_{\bar{q}}(\boldsymbol{x},t)$};
	\draw[|->,semithick] (1.5,1.5) -- (2.99,1.5) node[pos=0.2,anchor=north] {$\tilde{f}_q(\boldsymbol{x}-\boldsymbol{c}_q\Delta t, t)$};
	\draw[darkgray,|->,semithick] (4.5,1.55) -- (3.01,1.55) node[pos=0.48,anchor=south] {$f_{\bar{q}}(\boldsymbol{x}, t+\Delta t)$};
	\draw [decorate,decoration={brace,amplitude=4pt}] (4.51,1.5) -- (5.5,1.5) node [pos=0.5,anchor=south] {$\delta_q$};
	\draw[->,thick] (5.5,1.5) -- (6.6,2.0) node[pos=0.7,anchor=north west] {$\boldsymbol{v}_b$};
	\node[above] at (5.5,1.5) {$\boldsymbol{x}_b$};
	\fill (1.5,1.5) circle (2pt);
	\fill (4.5,1.5) circle (2pt);
	\fill (7.5,1.5) circle (2pt);
	\fill (5.5,1.5) circle (2pt);
	
	\draw[fill=lightgray,thin] (9.5,2.5) rectangle (10,3);
	\node[right] at (10,2.75) {\textit{solid}};
	\draw[pattern = custom north west lines, hatchspread=5pt, hatchcolor=lightgray, thin] (9.5,1.75) rectangle (10,2.25);
	\node[right] at (10,2) {\textit{fluid, near boundary}};
	\draw[thin] (9.5,1) rectangle (10,1.5);
	\node[right] at (10,1.25) {\textit{fluid}};
	\draw[blue,thick] (9.5,0.25) -- (10,0.25);
	\node[right] at (10,0.25) {particle surface};
	\end{tikzpicture}
	\caption{Sketch of the particle mapping and the boundary treatment according to the CLI boundary scheme from Eq.~\eqref{eq:MEM_CLI}.}
	\label{fig:BCsketch}
\end{figure}

A no-slip boundary condition is applied along the surface of the particles which is given by the central linear (CLI) scheme as \cite{ginzburg_two-relaxation-time_2008}
\begin{equation}
f_{\bar{q}}(\boldsymbol{x},t+\Delta t) = \tilde{f}_q(\boldsymbol{x},t) + \tfrac{1-2\delta_q}{1+2\delta_q} \left(\tilde{f}_q(\boldsymbol{x}-\boldsymbol{c}_q\Delta t, t) - \tilde{f}_{\bar{q}}(\boldsymbol{x}, t)\right)-\tfrac{12\, w_q\rho_0}{1+2\delta_q} \boldsymbol{v}_b \cdot \boldsymbol{c}_q, \label{eq:MEM_CLI}
\end{equation}
where the boundary velocity $\boldsymbol{v}_b = \boldsymbol{V}_p + \boldsymbol{W}_p \times (\boldsymbol{x}_b - \boldsymbol{X}_p)$ is the particle's velocity at the position $\boldsymbol{x}_b$.
The interpolation weight $\delta_q$ is the distance between the cell center $\boldsymbol{x}$ and the actual surface position $\boldsymbol{x}_b = \boldsymbol{x} + \delta_q \boldsymbol{c}_q$, normalized by the distance of the cell centers, i.e. $\|\boldsymbol{c}_q\Delta t\|$.
For spherical particles, it can be determined analytically as the result of a ray-sphere intersection problem.
A sketch of the boundary handling procedure is provided in Figure~\ref{fig:BCsketch}.

As a result of this boundary treatment, momentum is transferred from the fluid onto the submerged particle along the boundary link $q$, given as \cite{wen_galilean_2014}:
\begin{equation}
\boldsymbol{F}_q(\boldsymbol{x}_b, t) = \big(\boldsymbol{c}_q - \boldsymbol{v}_b\big) \tilde{f}_q(\boldsymbol{x},t) - \big(\boldsymbol{c}_{\bar{q}} - \boldsymbol{v}_b\big) f_{\bar{q}}(\boldsymbol{x}, t+\Delta t). \label{eq:MEM_Force}
\end{equation}

By summing up these contributions over all boundary links from all \textit{near-boundary} cells (NB) next to a single particle, the total hydrodynamic force and torque are determined as
\begin{align}
\boldsymbol{F}_p^{\text{hyd}} & = \sum_{NB} \sum_q \boldsymbol{F}_q(\boldsymbol{x}_b)\ , \label{eq:MEM_TotForce}\\
\boldsymbol{T}_p^{\text{hyd}} & = \sum_{NB} \sum_q (\boldsymbol{x}_b - \boldsymbol{X}_p) \times \boldsymbol{F}_q(\boldsymbol{x}_b), \label{eq:MEM_TotTorque}
\end{align}
respectively.

Finally, due to the explicit mapping of the particles into the domain, cells that convert from \textit{solid} to \textit{fluid} due to moving particles need to restore PDF information before the simulation can continue.
Here, we initialize the PDFs of such a transformed cell with its equilibrium values, Eq.~\eqref{eq:LBM_EQ}, using the particle's surface velocity and a spatially averaged density \cite{RETTINGER201774}.

\subsection{Parallelization}
\label{sec:domain_decomposition}

The direct numerical simulations of complex particulate flow scenarios can exhibit enormous computational costs that can only be tackled by parallel execution on a large number of processes.
Such large computers are designed with a distributed memory so that typically the message passing interface (MPI) is used to implement the parallel algorithms.
We here employ the open-source \textsc{waLBerla} framework \cite{walberla,Godenschwager2013} where all the above mentioned algorithms are implemented.
It partitions the computational domain into \textit{blocks}.
They contain all the data for this subdomain and are then assigned to the available MPI processes to distribute the workload, see Figure~\ref{fig:domain_part_blocks}.
This requires the introduction of synchronization mechanisms for the fluid as well as for the particle simulation which both use the same domain partitioning as a result of the close coupling within the same framework.

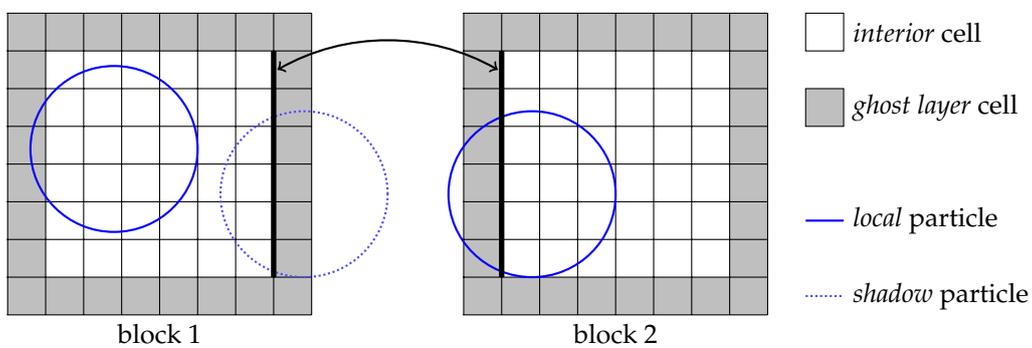
\begin{figure}[b]
	\centering
	\begin{tikzpicture}
	\fill[lightgray] (1,0.5) rectangle (5,4.5);
	\fill[white] (1.5,1.0) rectangle (4.5,4.0);
	\draw[step=0.5,black] (0.99,0.5) grid (5,4.5);
	\node[below] at (3,0.5) {block 1};
	
	\fill[lightgray] (7.0,0.5) rectangle (11,4.5);
	\fill[white] (7.5,1.0) rectangle (10.5,4.0);
	\draw[step=0.5,black] (6.99,0.5) grid (11,4.5);
	\node[below] at (9,0.5) {block 2};
	
	\draw[blue,thick] (7.9,2.1) circle (1.1);
	\draw[blue!70!white,thick,densely dotted] (4.9,2.1) circle (1.1);
	\draw[blue,thick] (2.4,2.7) circle (1.1);
	
	\draw[fill=white,thin] (11.5,4) rectangle (12,4.5);
	\node[right] at (12,4.25) {\textit{interior} cell};
	\draw[fill=lightgray,thin] (11.5,3) rectangle (12,3.5);
	\node[right] at (12,3.25) {\textit{ghost layer} cell};
	
	\draw[blue,thick] (11.5,1.75) -- (12,1.75);
	\node[right] at (12,1.75) {\textit{local} particle};
	\draw[blue!70!white,thick,densely dotted] (11.5,0.75) -- (12,0.75);
	\node[right] at (12,0.75) {\textit{shadow} particle};
	
	\draw[black,line width = 2] (4.5,1) -- (4.5,4);
	\draw[black,line width = 2] (7.5,1) -- (7.5,4);
	\draw[black,<->,thick] (7.45,3.75) arc (60:120:2.9);
	\end{tikzpicture}
	\caption{Domain partitioning into blocks. The cell-based fluid simulation requires a ghost layer for synchronization. The particle simulation uses shadow particles. The two bold vertical lines denote the common face of both blocks.}
	\label{fig:DomainDecomp}
\end{figure}

The sketch in Figure~\ref{fig:DomainDecomp} shows the applied techniques for two neighboring blocks.
Each block is subdivided into uniform cells that are required for the fluid simulation by the lattice Boltzmann method, Sec.~\ref{sec:LBM}.
To be able to carry out the streaming step, Eq.~\eqref{eq:LBM_Stream}, across block borders, an extra layer of cells, a so called \textit{ghost layer}, has to be introduced around each block.
They contain the corresponding PDFs from the neighboring blocks that are to be streamed to cells inside the block at hand and thus have to be communicated beforehand.
Similarly, the particle simulation defines \textit{local} and \textit{shadow} particles on a block \cite{Iglberger2009}. 
The former are particles whose center of mass are contained inside this block.
To check for particle-particle collisions, however, all possible collision partners have to be known on a block.
This necessitates that copies of all particles that intersect with the block, with the center of mass being contained inside another block, are present on that block as well.
These copies are called \textit{shadow} particles.

More details on the block-based domain partitioning and the communication routines can be found in \cite{Godenschwager2013, Schornbaum2016, Rettinger2017ISC, Schornbaum2018}.

\subsection{Complete Algorithm}
\label{sec:complete_algo}

By combining the different components presented in the previous sections, we obtain the complete algorithm for coupled fluid-particle simulations with geometrically fully resolved particles, shown in Algorithm~\ref{alg:LBM_MEM}.
Since the treatment of the boundary, Eq.~\eqref{eq:MEM_CLI}, and the link-based computation of the interaction force, Eq.~\eqref{eq:MEM_Force}, require the same information, we fuse them to a single step in line~\ref{alg:LBM_MEM:BH} of the algorithm.
Additionally, we note that the time step of the rigid body has to be chosen smaller than the one of the fluid simulation to obtain an accurate collision behavior \cite{BIEGERT2017105}. 
Therefore, we introduce a sub cycling loop for the rigid body solver in which we subdivide the time step into $S$ sub steps in which the hydrodynamic force acting on the particles is kept constant, see line~\ref{alg:LBM_MEM:SubCycle} of Algorithm~\ref{alg:LBM_MEM}.

\begin{algorithm}
	\DontPrintSemicolon
	\For{each time step $t$}{
		(Re-)map particle into fluid domain.\label{alg:LBM_MEM:Map}\;
		Reconstruct PDFs of converted cells if necessary.\label{alg:LBM_MEM:Recon}\;
		Perform LBM collision step, Eq.~\eqref{eq:LBM_Collide}.\label{alg:LBM_MEM:Collide}\;
		Communicate PDF ghost layers.\label{alg:LBM_MEM:ComPDF}\;
		Apply boundary conditions, Eq.~\eqref{eq:MEM_CLI} and compute hydrodynamic forces on particles, Eq.~\eqref{eq:MEM_Force}.\label{alg:LBM_MEM:BH}\;
		Perform LBM stream step, Eq.~\eqref{eq:LBM_Stream}.\label{alg:LBM_MEM:Stream}\;
		\For{each rigid body solver subcycle $S$\label{alg:LBM_MEM:SubCycle}} {
		Perform rigid body solver step (collision detection and resolution, force synchronization, time integration, position and velocity synchronization), Sec.~\ref{sec:NSGD}.\label{alg:LBM_MEM:RB}\;
	}
	}
	\caption{Coupled algorithm for particulate flow simulations.}
	\label{alg:LBM_MEM}
\end{algorithm}

\section{Development and Calibration of a Workload Estimator}
\label{sec:load_determination}

\subsection{Description}
The algorithm outlined in Algorithm~\ref{alg:LBM_MEM} is a typical representative for a simulation procedure to solve multiphysics problems.
One time step consists of contributions from the individual physical components, here fluid flow, Sec.~\ref{sec:LBM}, and rigid body dynamics, Sec.~\ref{sec:NSGD}, as well as coupling algorithms for their interaction.
Depending on the physical setup at hand, each of these parts generates a different workload within a block, used for the partitioning of the simulation domain, see Sec.~\ref{sec:domain_decomposition}.
Taking particulate flows as an example, blocks with dense particle packings result in longer compute times, i.e. larger workloads, for the rigid body simulation in comparison to blocks without any particles.
Additionally, in many physical problems, the location of these areas, and consequently the workload per block, changes over time.
Since the simulation gets synchronized via communications in each time step, unbalanced workload distributions result in waiting times for the processes that own blocks with smaller workloads.
These idle times reduce the overall efficiency of the simulation and result in a longer time to solution than ideally achievable.
For successful load balancing, it is thus important to first estimate the workload per block.
This is resembled by a scalar quantity and we denote it as \textit{weight} of the block, see Figure~\ref{fig:domain_part_blocks}.
Once the weights of all blocks are known, a load distribution algorithm can then be used to reassign the blocks to the processes to establish a balanced state. 

In this section, we present our approach to develop a load estimator for particulate flow simulations with Algorithm~\ref{alg:LBM_MEM}.
The goal is to find a function that is able to reliably predict the workload of a block, depending on local quantities that describe the state of this block.
The workload is here given as the runtime of a single time step in this simulation, excluding communication routines.
For that reason, simulations of a characteristic setup have to be carried out. 
During these simulations, the block local quantities and the corresponding runtimes of the different parts of the algorithm are continuously evaluated and stored.
Based on this data, function fitting is applied to finally obtain a estimator function that can be used in all subsequent simulations to predict the workload for a block.

The simulation setup hast to be chosen such that different cases which can be encountered in a real application also occur in the simulation.
This means that blocks with no particles and dilute as well as densely packed particles should be included in the setup and thus in the data.
Only then, the function that is to be fitted to these measurements can later on predict these cases accurately enough.
Specifically, we execute several simulations to obtain separate time measurements of the major parts of  Algorithm~\ref{alg:LBM_MEM} and vary some of the important parameters to increase the amount of data and the reliability.
The block local quantities that are available in our fluid-particle simulation are given in Table~\ref{tab:quantities} and their definition can be obtained from Figures~\ref{fig:BCsketch} and \ref{fig:DomainDecomp}.
For function fitting, we first have to determine the form of the functions that map these quantities to the measured runtime.
For each part, this is achieved by analyzing the structure of the individual algorithm.
In a calibration step, those functions are then fitted to the timing measurements of the separate parts based on the block local quantities to determine the functions' coefficients.
Finally, those functions are combined to obtain a complete estimator for the workload per block.

We note, that this procedure must be carried out only once as a preprocessing step for all upcoming simulations.
Once the load estimator is found, it is simply applied, i.e. the fitted function is evaluated, in the simulations whenever load balancing is carried out.
This procedure also only involves simulations of small systems which keeps the computational effort small.

\begin{table}[t]
		\caption{Quantities that describe the state of a block.}
		\label{tab:quantities}
	\centering
	\begin{tabular}{c|l}
		\toprule
		variable & description \\
		\midrule
		$C$ & total number of cells on a block\\
		$F$ & number of cells flagged as \textit{fluid}\\ 
		$B$ & number of cells flagged as \textit{near boundary}\\
		$P_L$ & number of local particles \\ 
		$P_S$ & number of shadow particles \\
		$K$ & number of contacts between rigid bodies \\
		$S$ & number of sub cycles of the rigid body simulation\\
		\bottomrule
	\end{tabular}
\end{table}

\subsection{Workload Contributions}
\label{sec:WL_contributions}

In the following, we establish the form of the functions we use to fit the measured times based on the local quantities from Table~\ref{tab:quantities}.
The contributions are mainly gathered from the structure of the respective implementation, which are briefly outlined for each part.

\subsubsection{LBM module}

The simulation of the fluid flow is carried out by the LBM, which consists of the collision (line \ref{alg:LBM_MEM:Collide} of Algorithm~\ref{alg:LBM_MEM}) and the stream (line \ref{alg:LBM_MEM:Stream}) step.
These are only carried out for cells marked as \textit{fluid} and the pseudocode is given in Algorithms~\ref{alg:LBM_collide} and \ref{alg:LBM_stream}, respectively.
The resulting workload heavily depends on the applied collision operator and the actual implementation \cite{WELLEIN2006910,WITTMANN2018582} but it will generally be mainly determined by the number of cells $C$ and the number of fluid cells $F$.
Thus, a function that represents the workload generated by the two LBM steps is given by
\begin{equation}
\text{WL}_{\text{LBM}} = a_{1,\text{LBM}} C + a_{2,\text{LBM}} F  + a_{3,\text{LBM}} \label{eq:WL_LBM}
\end{equation}

\begin{algorithm}[h]
	\DontPrintSemicolon
	\For{each cell $c$}{
		\If{$c$ is a fluid cell}{
		Perform LBM collision step.\;
	}
}
	\caption{Pseudocode for LBM (collision).}
	\label{alg:LBM_collide}
\end{algorithm}

\begin{algorithm}[h]
	\DontPrintSemicolon
	\For{each cell $c$}{
		\If{$c$ is a fluid cell}{
		Perform LBM stream step.\;
	}
	}
	\caption{Pseudocode for LBM (stream).}
	\label{alg:LBM_stream}
\end{algorithm}

\subsubsection{Boundary handling module}

The boundary handling procedure for LBM, line \ref{alg:LBM_MEM:BH} of Algorithm~\ref{alg:LBM_MEM}, applies the chosen condition for each near boundary cell.
More specifically, it does so for each fluid-boundary link of the near boundary cell
As can be seen in the pseudocode of Algorithm~\ref{alg:BH}, the resulting workload will thus be related to the number of these links.
However, evaluating this number for each block is computationally more costly than simply counting the number of near boundary cells and the workload per link depends on the specific boundary condition, which we here do not distinguish.
We therefore simply use a function of the form
\begin{equation}
\text{WL}_{\text{BH}} = a_{1,\text{BH}} C + a_{2,\text{BH}} NB + a_{3,\text{BH}}, \label{eq:WL_BH}
\end{equation}
where we additionally included $C$ to represent the outer loop.

\begin{algorithm}
	\DontPrintSemicolon
	\For{each cell $c$}{
	\If{$c$ is a near boundary cell}{
	\For{each neighboring boundary cell}{
	Perform boundary handling according to applied boundary condition.\;
}
}
}
\caption{Pseudocode for boundary handling.}
\label{alg:BH}
\end{algorithm}

\subsubsection{Particle mapping module}

Mapping the particles into the domain by marking the contained cells as boundary cells is an important part of the applied fluid-particle coupling algorithm, see line \ref{alg:LBM_MEM:Map} of Algorithm~\ref{alg:LBM_MEM}.
The pseudocode is outlined in Algorithm~\ref{alg:Map}.
The resulting workload thus depends on the number of particles, local and shadow, as well as the extent of their axis-aligned bounding box.
Since the size of this box depends on the diameter of the particle, this information is not readily available.
Instead, we can analyze the result of the mapping by including the number of non-fluid cells as a kind of block local solid volume fraction.
Since shadow particles are only partially present on the block they will generate a smaller workload than local particle and we should thus distinguish workload contributions originating from local and shadow particles in our function.
We therefore propose the following function:
\begin{equation}
\text{WL}_{\text{Coup1}} = a_{1,\text{Coup1}} C + a_{2,\text{Coup1}} F + a_{3,\text{Coup1}} P_L + a_{4,\text{Coup1}} P_S + a_{5,\text{Coup1}}
\label{eq:WL_Coup1}
\end{equation}

\begin{algorithm}
	\DontPrintSemicolon
		\For{each particle $p$}{
		Obtain an axis-aligned bounding box (AABB) which fully contains the particle.\;
		\For{each cell $c$ in this AABB}{
		\If{cell $c$ is contained in particle $p$}{
		Set flag to boundary (solid) flag.\;
}
}
}
	\caption{Pseudocode for particle mapping (Coup1).}
	\label{alg:Map}
\end{algorithm}

\subsubsection{PDF reconstruction module}

The second part of the coupling algorithm, line \ref{alg:LBM_MEM:Recon} in Algorithm~\ref{alg:LBM_MEM}, reconstructs the PDF values in cells that have changed from being solid to being fluid due to the motion of the corresponding particle, see Algorithm~\ref{alg:Recon}.
The generated workload thus depends on the position, orientation and velocity of the individual particles and is therefore difficult to predict in general.
Assuming that chances for these cell state changes are higher if there are more particles around and the solid volume fraction is larger, we estimate its workload by:

\begin{equation}
\text{WL}_{\text{Coup2}} = a_{1,\text{Coup2}} C + a_{2,\text{Coup2}} F + a_{3,\text{Coup2}} P_L + a_{4,\text{Coup2}} P_S + a_{5,\text{Coup2}}
\label{eq:WL_Coup2}
\end{equation}

\begin{algorithm}
	\DontPrintSemicolon
\For{each cell $c$}{
	\If{$c$ has switched from solid to fluid cell}{
		Reconstruct PDFs in $c$.\;
	}
}
\For{each cell $c$}{
	\If{$c$ has switched from solid to fluid cell}{
		Set fluid flag in $c$.\;
	}
}
	\caption{Pseudocode for PDF reconstruction (Coup2).}
	\label{alg:Recon}
\end{algorithm}

\subsubsection{Rigid body simulation module}

The rigid body simulation part, line \ref{alg:LBM_MEM:RB} of Algorithm~\ref{alg:LBM_MEM}, consists of several components which for simplicity will all be included in a single function.
Algorithm~\ref{alg:RB} outlines these different sub steps \cite{Preclik2015}.
The first part, contact detection, is typically of squared complexity since all particles have to be checked against all other particles to find possible contacts.
Our implementation makes use of hash grids for optimizing this routine to obtain linear complexity.
This, however, is only activated if a reasonable number of particles is present on a block since otherwise the computational overhead renders it less efficient.
In a second step, the contact resolution treats each contact according to the applied collision model in order to resolve the overlaps between particles.
Determining the needed information for these two steps is simple for spherical particles but can become complex and costly for other shapes.
The last step uses a time integrator scheme to update the local particles' position and velocity.
In the simulation algorithm, the whole rigid body algorithm is embedded into a loop over $S$ number of sub cycles, see line \ref{alg:LBM_MEM:SubCycle} of Algorithm~\ref{alg:LBM_MEM}.
Combining these steps, we use the following function to estimate the workload:

\begin{equation}
\text{WL}_{\text{RB}} = S \left( a_{1,\text{RB}} (P_L + P_S)^2 + a_{2,\text{RB}}P_L + a_{3,\text{RB}}P_S + a_{4,\text{RB}} K + a_{5,\text{RB}}\right)
\label{eq:WL_RB}
\end{equation}

\begin{algorithm}
	\DontPrintSemicolon
	\For{each particle $p$}{
		\If{$p$ is in contact with another particle}{
		Generate a contact pair.\;
}
}
		\For{each contact pair $k$}{
		Compute reaction to resolve contact.\;
}
		\For{each particle $p$}{
		Integrate $p$ in time using the acting forces.\;
}
	\caption{Pseudocode for rigid body simulation.}
	\label{alg:RB}
\end{algorithm}

\subsubsection{Total workload estimator}

Since, in the general case, only a single weight per block has to be provided, the aforementioned contributions must be combined in order to obtain a estimator function for the total workload per block.
A natural and simple choice is to add up all individual functions, Eqs.~\eqref{eq:WL_LBM}--\eqref{eq:WL_RB}, effectively combing the coefficients of the block quantities:
\begin{equation}
\text{WL}_\text{tot} = \text{WL}_{\text{LBM}} + \text{WL}_{\text{BH}} + \text{WL}_{\text{Coup1}} +\text{WL}_{\text{Coup2}} + \text{WL}_{\text{RB}}.
\label{eq:WL_tot}
\end{equation}

\subsection{Simulation Setup}
\label{sec:load_determination_setup}

\begin{figure}
	\centering
	\begin{subfigure}[t]{0.32\textwidth}
		\centering
		\includegraphics[width=\textwidth]{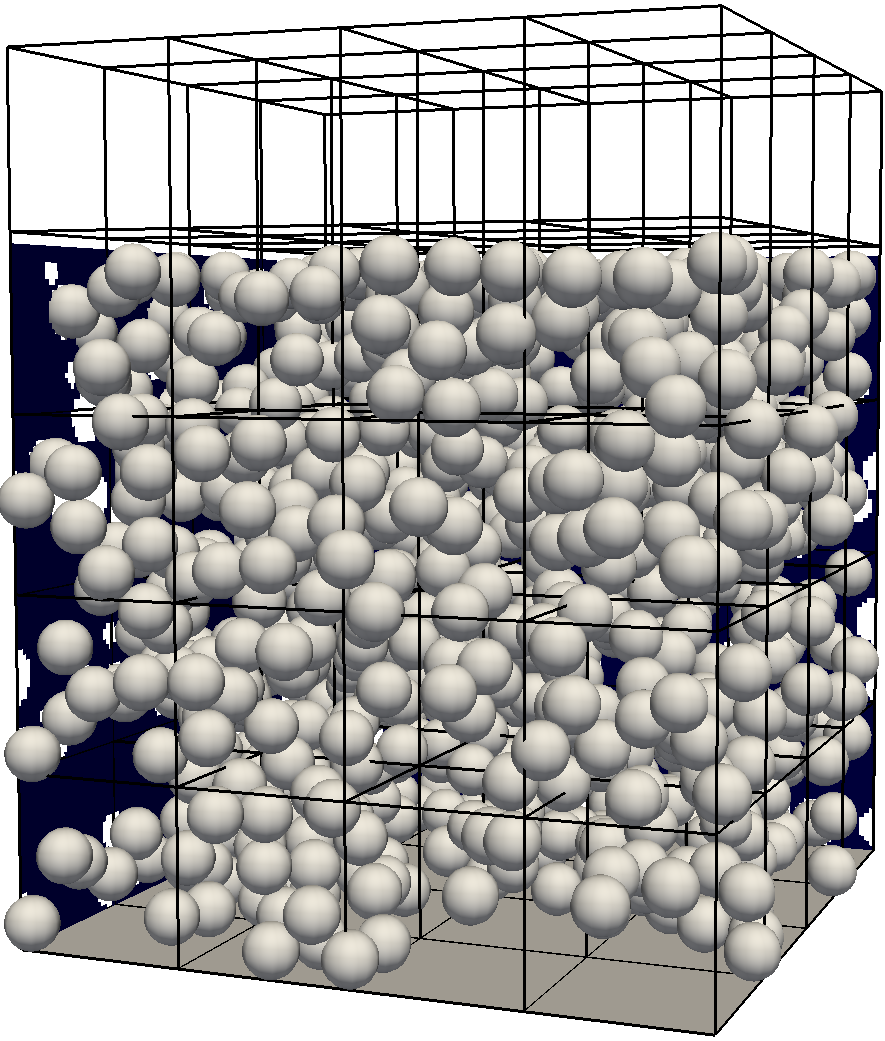}
		\caption{Initially random particle distribution ($t^*=0$). }
	\end{subfigure}
	~
	\begin{subfigure}[t]{0.32\textwidth}
		\centering
		\includegraphics[width=\textwidth]{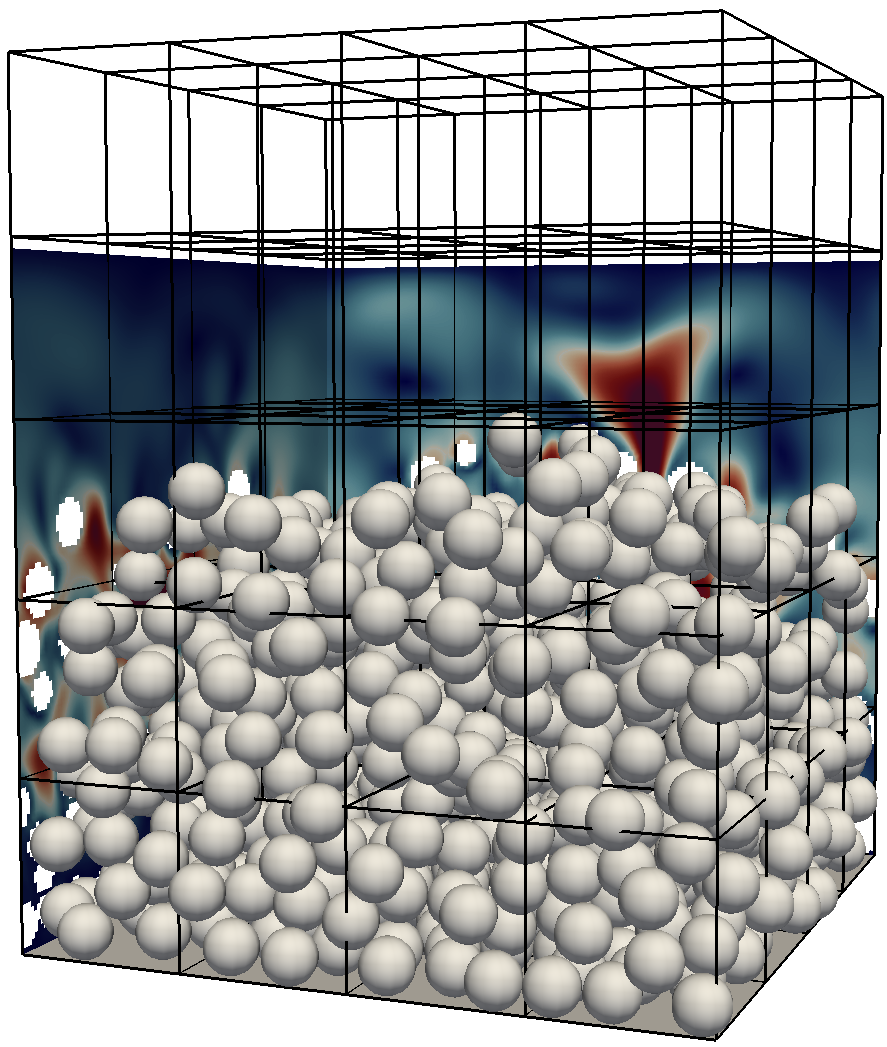}
		\caption{Ongoing settling ($t^*=1.3$).}
	\end{subfigure}
	~
	\begin{subfigure}[t]{0.32\textwidth}
		\centering
		\includegraphics[width=\textwidth]{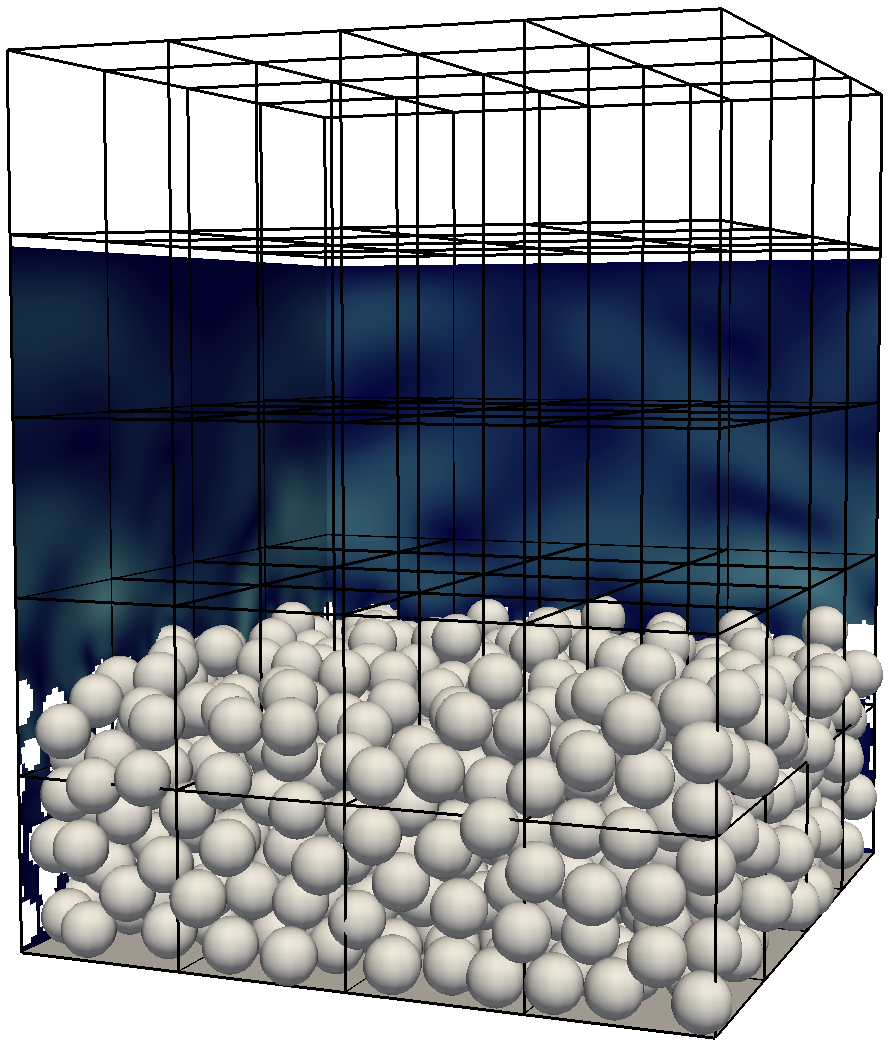}
		\caption{Final, completely settled state ($t^*=2.5$).}
	\end{subfigure}
	\caption{Workload evaluation simulation with $b_s/\Delta x=32$ and $D/\Delta x=10$. Black boxes show the $4\times4\times5$ blocks used for domain partitioning. The magnitude of the fluid velocity is shown in two slices at the back of the domain.}
	\label{fig:WL_sim}
\end{figure}

After having identified and defined the workload functions for the different modules of the algorithm, we now carry out multiple simulations of a characteristic setup to obtain timing information that are then used to fit the functions and determine the coefficients.
To obtain a generally applicable estimator, the setup should be designed such that it contains various cases that are also encountered in typical applications involving particulate flows.
Those are e.g. bounding walls, densely packed areas with many inter-particle collisions as well as dilute regions with only few or even no particles.
Additionally, due to complex bounding geometries, some parts of the computational domain could be completely excluded from the simulation as neither fluid nor particles can enter these regions. 
We therefore chose a horizontally periodic setup with an initially random distribution of particles.
Those particles are then set into a settling motion due to the acting gravity and then pack at the bottom plane until all particles have settled.
Visualizations of the initial, intermediate and final state of the simulations with spherical particles can be seen in Figure~\ref{fig:WL_sim}.

The characteristic parameters of this setup are the Galileo number, $Ga =  u_s D / \nu = 30$, with a characteristic settling velocity $u_s = \sqrt{ (\rho_p/\rho_f -1 ) g D}$, the diameter $D$ and kinematic viscosity $\nu$.
The particles have a density ratio of $\rho_p / \rho_f = 2.5$ and are subjected to a gravitational acceleration $g$ in vertical direction, resulting in gravitational and buoyancy forces.
The domain size is $L_x \times L_y  \times L_z$ with $L_x = 4 b_s$, $L_y=4 b_s$ and $L_z = 5 b_s - o_t$.
Here, $b_s$ is the block size and we introduce a constant offset of the top wall $o_t = 1.05 b_s$.
The global solid volume fraction of the domain is $0.2$.
The simulation is run for $2.5$ unitless time steps, where one time step is $T = L_z / u_s$.
As the definition of the domain size already suggests, we subdivide the whole domain into $4\times4\times5$ blocks of size $b_s \times b_s \times b_s$.

As a result, we obtain five vertical layers of blocks with different characteristics throughout the simulation.
The bottom layer experiences a continuous increase in the number of particles until it is densely packed.
The second layer features the interface between the particle packing and the bulk fluid region at the end of the simulation.
Blocks of the third layer are traversed by all upper particles during the settling phase and end up without any particles or boundaries.
Similarly, the fourth layer ends up with no particles but, due to the offset in the top wall, with a constant number of boundary and near-boundary cells.
The blocks of the topmost row contain neither fluid cells nor particles throughout the complete simulation since they are completely overlapped by the top wall.

Those simulations are carried out for different typical block sizes $b_s/\Delta x = \{24, 32, 48, 64\}$ and diameters $D/\Delta x = \{10,20\}$ to obtain a large enough variance in the samples.
Each simulation is executed in parallel on $80$ processes of the SuperMUC Haswell cluster at the LRZ such that each process is assigned one block.
This allows to obtain separate measurements for each block with $2000$ samples each.
In total, this results in around $1.3\times10^6$ data points that will be used for the function fitting.
In the first $50$ time steps, no measurements are taken to exclude possible warm-up effects of the hardware.
Furthermore, we make use of thread pinning provided by the LIKWID tool \cite{likwidWebsite,likwid} to obtain reliable measurements.
Each sample consists of the current values for the block-local quantities from Table~\ref{tab:quantities}, where $S=10$ is kept in all simulations, and additionally timing measurements of the five algorithm parts from Sec.~\ref{sec:WL_contributions}, i.e. $m_\text{LBM}, m_\text{BH}, m_\text{Coup1}, m_\text{Coup2}$ and $m_\text{RB}$.

\subsection{Results of the Calibration}
\label{sec:WL_Results}

\begin{figure}[t]
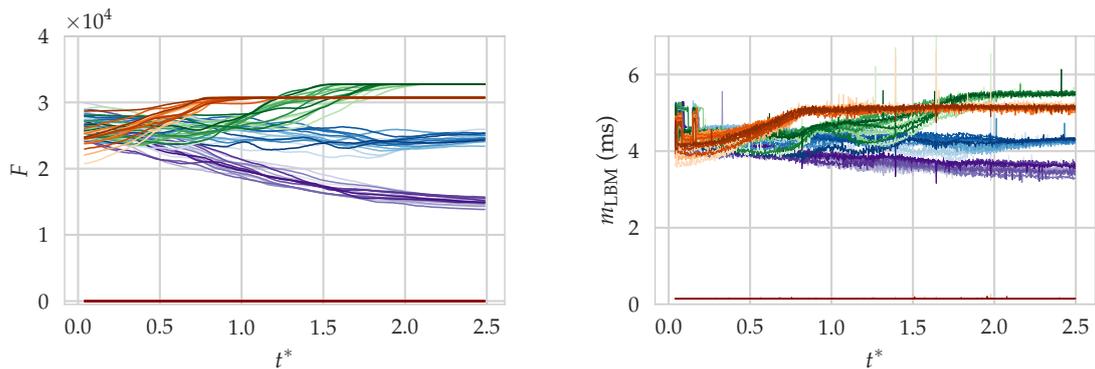

\centering
\begin{subfigure}[t]{0.48\textwidth}
	\centering
	\input{figures/wl_F_paper.pgf}
\end{subfigure}
~
\begin{subfigure}[t]{0.48\textwidth}
	\centering
	\input{figures/wl_tLBM_paper.pgf}
\end{subfigure}
\caption{Number of fluid cells and time measurements (in ms) of the LBM part per block over time ($t^*=t/T$) of the workload evaluation simulation with $b_s/\Delta x=32$ and $D/\Delta x=10$. The coloring resembles the affiliation of the block to one of the five rows in the setup.}
\label{fig:WL_measurement}
\end{figure}

An example of the quantity evaluation and time measurement for the LBM module of the aforementioned simulation setup is shown in Figure~\ref{fig:WL_measurement}. 
From the temporal evolution of $F$, the curves can be matched to a block of one of the five rows.
Clearly, a correlation between $F$ and $m_\text{LBM}$ can be seen which is in agreement with the assumption made in Eq.~\eqref{eq:WL_LBM}.

\begin{figure}[t]
	\centering
	\input{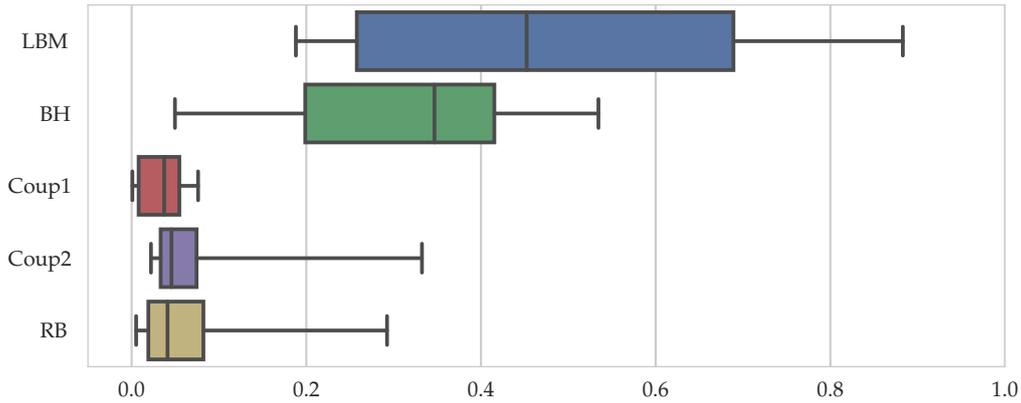}
	\caption{Box-and-whiskers plot of the proportion of part's runtime of total runtime, $m_{\text{X}}/m_{\text{tot}}$, for each sample.}
	\label{fig:boxplot_runtime}
\end{figure}

Before we evaluate the result of the calibration process of the workload estimator, we want to obtain insight into the proportion of the different parts of the total runtime, given by
\begin{equation} 
m_\text{tot} = m_\text{LBM} + m_\text{BH} + m_\text{Coup1} + m_\text{Coup2} + m_\text{RB}.
\label{eq:mtot}
\end{equation}

These proportions are different for each sample and are thus visualized in a compact format as a box-and-whiskers plot in Figure~\ref{fig:boxplot_runtime}, showing the median, the upper, and lower quartile values as a box, as well as whiskers that extend to the $5$ and $95$ percentiles.
It can be seen that the LBM and the BH part make up most of the workload with an average of $45$\% and $35$\%, respectively.
This is followed by the rigid body simulation and the coupling parts with median values below $5$\%.
The relatively large spread in the whiskers is introduced by the empty blocks of the fifth block row in the setup where total runtimes are very low and the different parts, except for Coup1, contribute similarly to it.
We can conclude that for spherical particles with the chosen coupling algorithm, it is most important to accurately predict the workload from the lattice Boltzmann and the boundary handling routines.

With all samples available, we make use of the curve fitting functionality provided by the Python library SciPy in a post-processing step to obtain the coefficients from Eqs.~\eqref{eq:WL_LBM}-\eqref{eq:WL_RB}.
The outcome is presented in Table~\ref{tab:fitted_coefs}.

\begin{table}
	\caption{Fitted values for the coefficients of Eqs.~\eqref{eq:WL_LBM}-\eqref{eq:WL_RB}.}
	\label{tab:fitted_coefs}
	\centering
	\begin{tabular}[t]{l|r}
		\toprule
	coefficient & value \\\midrule
$a_{1,\text{LBM}}$ & 9.99e-06\\
$a_{2,\text{LBM}}$ & 1.57e-04\\
$a_{3,\text{LBM}}$ & -8.23e-02\\
$a_{1,\text{BH}}$ & 6.65e-06\\
$a_{2,\text{BH}}$ & 7.06e-04\\
$a_{3,\text{BH}}$ & -1.09e-01\\
\bottomrule
\end{tabular}
~
	\begin{tabular}[t]{l|r}
		\toprule
		coefficient & value \\\midrule
$a_{1,\text{Coup1}}$ & 3.08e-06\\
$a_{2,\text{Coup1}}$ & 2.42e-07\\
$a_{3,\text{Coup1}}$ & 1.41e-02\\
$a_{4,\text{Coup1}}$ & 2.78e-02\\
$a_{5,\text{Coup1}}$ & -1.40e-01\\
\bottomrule
	\end{tabular}
	~
	\begin{tabular}[t]{l|r}
		\toprule
	coefficient &  value \\\midrule
$a_{1,\text{Coup2}}$ & 5.99e-06\\
$a_{2,\text{Coup2}}$ & 3.90e-06\\
$a_{3,\text{Coup2}}$ & -8.80e-03\\
$a_{4,\text{Coup2}}$ & 2.51e-02\\
$a_{5,\text{Coup2}}$ & -1.30e-01\\
\bottomrule
\end{tabular}
~
\begin{tabular}[t]{l|r}
	\toprule
	coefficient & value \\\midrule
$a_{1,\text{RB}}$ & 1.16e-06\\
$a_{2,\text{RB}}$ & 9.62e-04\\
$a_{3,\text{RB}}$ & 2.75e-04\\
$a_{4,\text{RB}}$ & 1.48e-03\\
$a_{5,\text{RB}}$ & 1.88e-02\\
\bottomrule
	\end{tabular}

\end{table}

\begin{figure}[t]
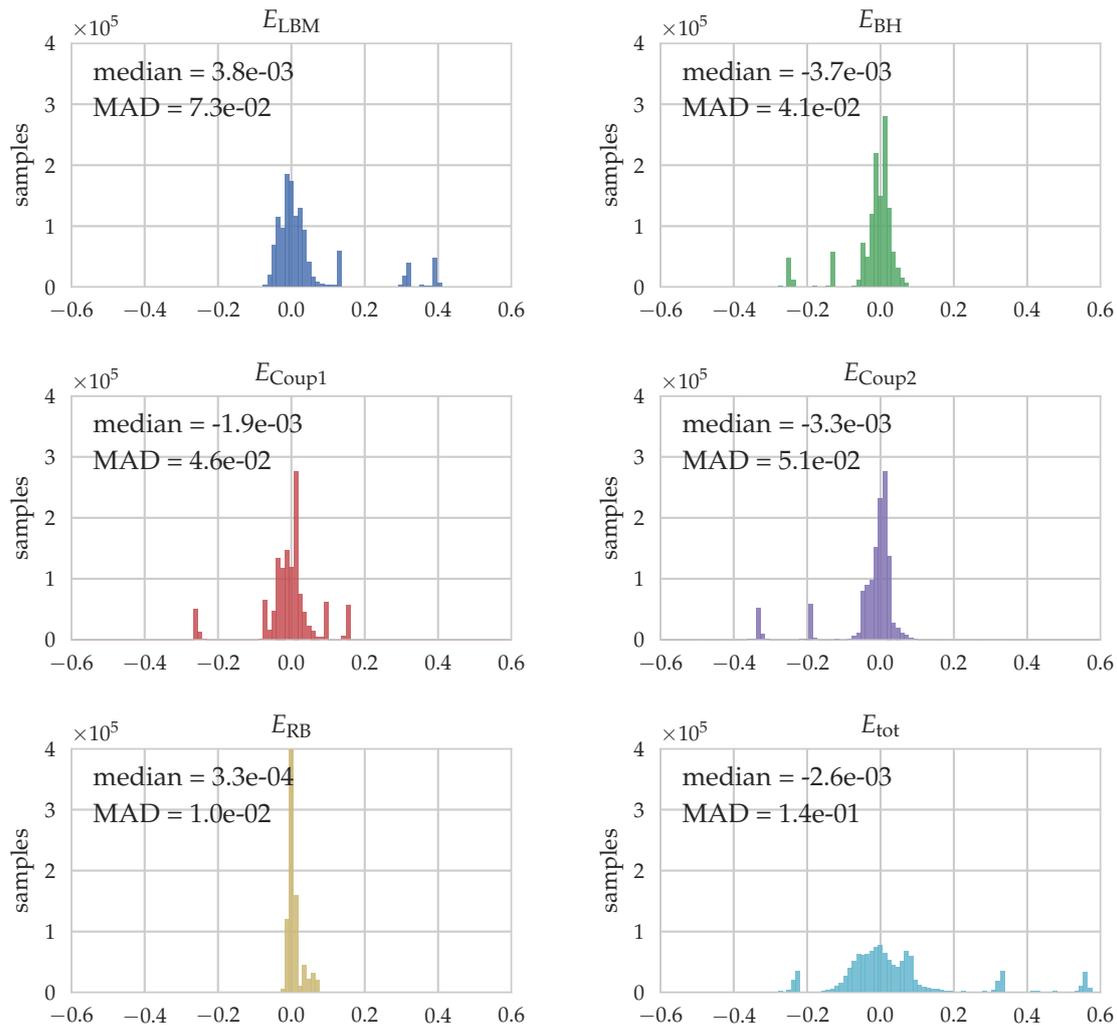

	\centering
	\begin{subfigure}[t]{0.48\textwidth}
		\centering
		\input{figures/wl_relTotDiffLBM_paper.pgf}
	\end{subfigure}
	~
	\begin{subfigure}[t]{0.48\textwidth}
		\centering
		\input{figures/wl_relTotDiffBH_paper.pgf}
	\end{subfigure}
	\begin{subfigure}[t]{0.48\textwidth}
		\centering
		\input{figures/wl_relTotDiffCoup1_paper.pgf}
	\end{subfigure}
	~
	\begin{subfigure}[t]{0.48\textwidth}
		\centering
		\input{figures/wl_relTotDiffCoup2_paper.pgf}
	\end{subfigure}
	\begin{subfigure}[t]{0.48\textwidth}
		\centering
		\input{figures/wl_relTotDiffRB_paper.pgf}
	\end{subfigure}
	~
	\begin{subfigure}[t]{0.48\textwidth}
		\centering
		\input{figures/wl_relTotDiffTot_paper.pgf}
	\end{subfigure}
	\caption{Relative errors $E_\text{X}$ of predicted workload for the five different parts and the total runtime.}
	\label{fig:WL_relTotDiff_eval}
\end{figure}

To quantify the quality of the fit for each part, various measures can potentially be evaluated.
One of them is the relative error of predicted workload by the five different fits for each individual sample, calculated as $E_\text{X} = (\text{WL}_\text{X} - m_\text{X})/m_\text{tot}$. 
It is visualized as histograms for the different parts and the total workload in Figure~\ref{fig:WL_relTotDiff_eval}.
Additionally, the median and the median absolute deviation (MAD), which is a robust measure of the statistical dispersion, are stated. 
Note that we calculate the relative error with respect to the total runtime instead of $m_\text{X}$ to avoid the division by almost zero for the small runtimes of Coup1 and RB, see Figure~\ref{fig:boxplot_runtime}.

For the LBM part, we obtain good predictions of the workload with a median close to zero. 
Outliers can be seen at some distinct error values which originate from the empty blocks where the prediction is generally worse.
A similar conclusion can be drawn for the BH part, where again the median is close to zero and some outliers can be found, this time on the negative side, i.e. due to underpredictions of the workload. 
These samples can again be attributed to those with empty blocks but also the ones of the third row which eventually end up with no boundary cells like the empty blocks.
The Coup1 part shows an acceptable median value but a relatively large MAD value due to the outliers at around $30\%$. 
These are again the empty blocks where the assumption we have formulated when establishing our fit function in Eq.~\eqref{eq:WL_Coup1}, i.e. the mapping should be related to the number of boundary cells as a measure of the solid volume fraction, breaks down as no particles are present.
Similarly, for Coup2 the median is close to zero but the MAD value is relatively large.
As discussed when setting up the equation for the fit, Eq.~\eqref{eq:WL_Coup2}, predicting the workload for this part is especially difficult as it can not be directly related to any of the available quantities.
This is then also the source of deviations from the predictions.
At last, the RB part is well-predicted in average and also the variance is small.
Considering the relative error for the total runtime, shown in the last histogram, the obtained estimator from Eq.~\eqref{eq:WL_tot} is able to predict the observed runtimes with good accuracy.
Outliers originating from the empty blocks of different block sizes where the fits do not work as good can be seen.
We also find that over $85\%$ of all samples are predicted with a relative error of less than $10\%$.

\subsection{Discussion}

In this section, we have constructed a specific workload estimator for coupled fluid-particle simulations with the methods from Sec.~\ref{sec:numerical_method}.
Here, the derivation of the specific form of the fit functions, Eqs.~\eqref{eq:WL_LBM}-\eqref{eq:WL_RB}, was purely driven by the algorithms themselves.
These functions can thus only be valid assumptions for the here applied algorithms.
As they calibration process of the coefficients relies on data from actual runtime measurements, the obtained coefficients for these functions also depend on the hardware on which the corresponding measurements were obtained.
However, it can be expected that they still remain good predictors if the hardware design is similar to the used one. 
In case of larger differences, like e.g. using a GPU instead of a CPU to run the simulation \cite{FEICHTINGER20151, Riesinger2017}, this might no longer be the case and the measurements and fits should be redone.
A possible improvement to overcome this drawback would be to also add hardware details, like cache sizes, clock frequency, etc., to the estimator and try to come up with a performance model for hardware-aware predictions.
Since this is a topic of active research, even for simple, stencil-based algorithms like LBM \cite{POHL2003,WITTMANN2018582}, this is not further pursued here.

Generally, the here reported values for the coefficients from Table~\ref{tab:fitted_coefs} should not be blindly reused in other programs even if the same algorithms and hardware are used, as also the actual implementation of the algorithms can have a large influence on the runtimes.
Nevertheless, by following the procedure presented in this section as a general recipe, workload estimators for many different applications, including pure CFD or complicated multiphysics setups, can be obtained.
They are then tailored and calibrated for the specific algorithms, hardware and implementation, and can be used in upcoming simulations of that kind to predict, and in a next step also distribute, the workload.

Besides the form of the fit functions, the simulation setup to generate the measurements has to be selected with care.
As reported in Sec.~\ref{sec:WL_Results}, in our case the predictions of the empty blocks were generally worse than for other, regular blocks as some assumptions broke down.
We still included them in the calibration procedure in order to obtain a general workload estimator and to avoid any special treatment of these empty blocks in our load estimation step.
However, if empty blocks do not appear in the simulations in which this workload estimator will then be applied, we suggest to remove them from the test scenario as well.
This improves the quality of the fits notably and a more accurate estimator can be obtained.

\section{Comparison of Load Distribution Methods}

\label{sec:comparison_load_distribution}

\subsection{Description}

Once the calibration process of the workload estimator has been executed, Eq.~\eqref{eq:WL_tot} with coefficients from Table~\ref{tab:fitted_coefs}, it can be applied in the subsequent simulations to assign a weight to each block, representing its workload.
As a second step, these weights are now used by load distribution methods to reassign the blocks to the processes dynamically throughout the simulation.
Besides the reduction of load imbalances, minimizing communication costs between the different processes and keeping the redistribution costs low are the key aspects of this step. 
We briefly review the different types of distribution methods in the next section.
Next, we set up a simulation that is different from the one used to deduce the load estimator in Sec.~\ref{sec:load_determination_setup}.
With this setup, we demonstrate the capabilities of our workload estimation approach and can compare the performance of different load distribution strategies.

\subsection{Load Distribution Algorithms}
\label{sec:load_dist_strategies}

\subsubsection{Space-filling curve}

Generally, a space-filling curve has the property that it covers the entire n-dimensional unit hypercube. 
Typically, these curves are constructed recursively from a sequence of piecewise linear continuous curves, following a specific construction pattern.
Different versions of such curves exist, e.g. the Peano, Hilbert, or Morton curve, as reviewed in \cite{bader2012space}.
As a practical result, a curve is obtained that connects all blocks of the computational domain and thus determines a one-dimensional ordering of these blocks.
For load distribution and while knowing the block weights as well as the total weight of all blocks, the blocks are assigned to the processes in a greedy manner.
This means that one traverses the curve and picks the blocks until the sum of these block weights is approximately equal to the total weight divided by the number of processes.
Due to the construction of these curves, the assigned blocks are usually geometrical neighbors and thus a reduction of the inter-process communication is implicitly achieved.
More information about space-filling curves can be found in \cite{bader2012space}.

\subsubsection{Diffusive distribution}

Load distribution via space-filling curves requires global information about the blocks and their weights.
For massively parallel simulations, this procedure poses an upper limit on the applicable number of processes as it inherently does not scale and also might require too much memory on a single process \cite{Schornbaum2018,eibl2018systematic}.
For these cases, diffusive techniques become the only feasible approach to distribute the load.
There, processes require only information about the load distribution in their direct neighborhood and will then try to even out possible imbalances by sending individual blocks to these neighboring processes.
The load gets distributed in a diffusive manner and thus several iterations are required to obtain good results.
However, a well-balanced load cannot be guaranteed and also the blocks on a process might get fragmented over time which increases communication costs.

\subsubsection{Graph partitioning}

The primary goal of these approaches is to partition unstructured graphs such that the edge cut is minimized, corresponding to a reduced communication between the processes.
This requires to determine weights for the edges that resemble the communication cost between neighboring blocks.
As an additional optimization constraint, weights for the vertices, i.e. here the blocks, can be provided in order to balance these as well.
Since graph partitioning is a common problem in various application fields, several algorithms are available \cite{HENDRICKSON2000485,karypisKumar1998, Buluc2016}.
The MPI-parallel library ParMETIS \cite{parmetis} is a commonly applied choice for this multi-objective optimization task.
It offers various algorithms to construct, improve or update graph partitions that can be further tuned by specifying additional parameters.
Furthermore, if provides functionalities to deal with multi constraint problems where several weights per block can be given to account for multiphysics applications \cite{Karypis1998,Schloegel2002}. 
Generally, it makes use of combinations of space-filling curves, multilevel algorithms and diffusive distributions to obtain a graph partitioning. 

Specifically, the required edge weights $EW$ between two vertices $v_1$ and $v_2$, i.e. here blocks with $b_s^3$ cells, are evaluated as follows to resemble the communication volume for the LBM part:
\begin{equation}
\text{EW}(v_1, v_ 2)= 
\begin{cases}
b_s^2, \text{ if } v_1 \text{ and } v_2 \text{ have common face},\\
b_s, \text{ if } v_1 \text{ and } v_2 \text{ have common edge},\\
1, \text{ if } v_1 \text{ and } v_2 \text{ have common corner},\\
0, \text{ else}.\\
\end{cases}
\label{eq:edge_weights}
\end{equation}

ParMETIS offers various algorithms and tuning parameters.
We restrict ourselves to the algorithms \texttt{PartGeomKway} and \texttt{AdaptiveRepart}.
The latter one is chosen since it is supposedly particularly well-suited for simulations with adaptive grid refinement, which is one natural use case of load balancing in general.
The proposed default values for the load imbalance tolerance, \texttt{ubvec} $=1.05$, and the ratio between inter-process communication time compared to data redistribution times, \texttt{itr} $=1000$, are used.

\subsection{Simulation Setup}

\begin{figure}
	\centering
	\begin{subfigure}[t]{0.32\textwidth}
		\centering
		\includegraphics[width=\textwidth]{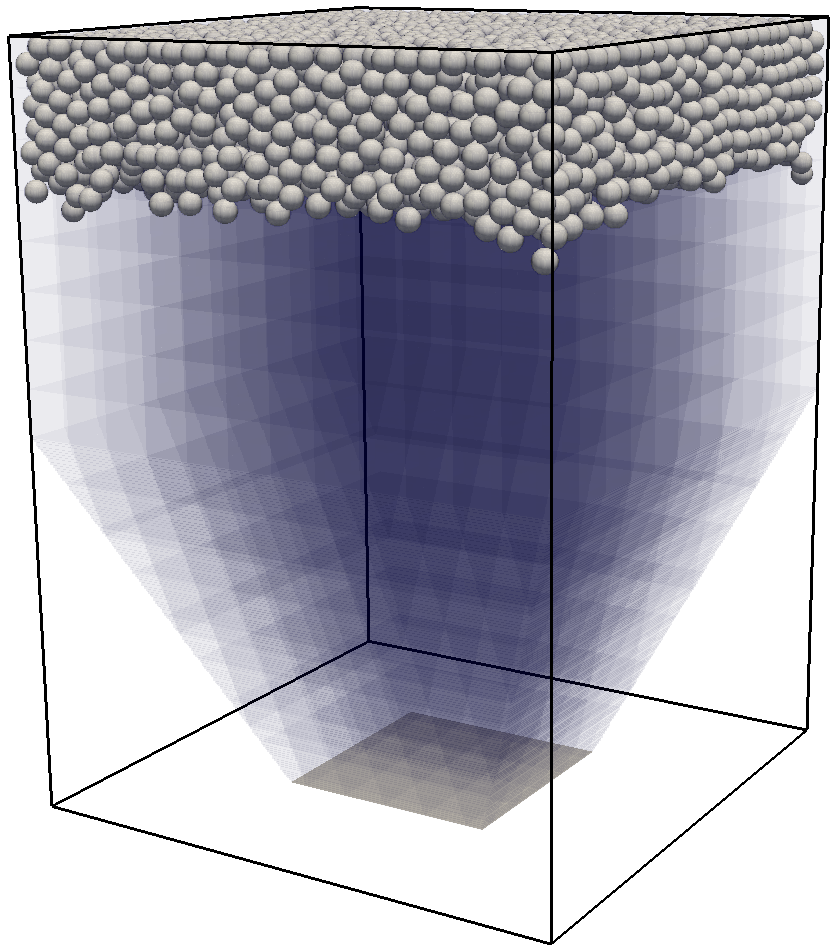}
		\caption{Initial state with block structure.}
	\end{subfigure}
	~
	\begin{subfigure}[t]{0.32\textwidth}
		\centering
		\includegraphics[width=\textwidth]{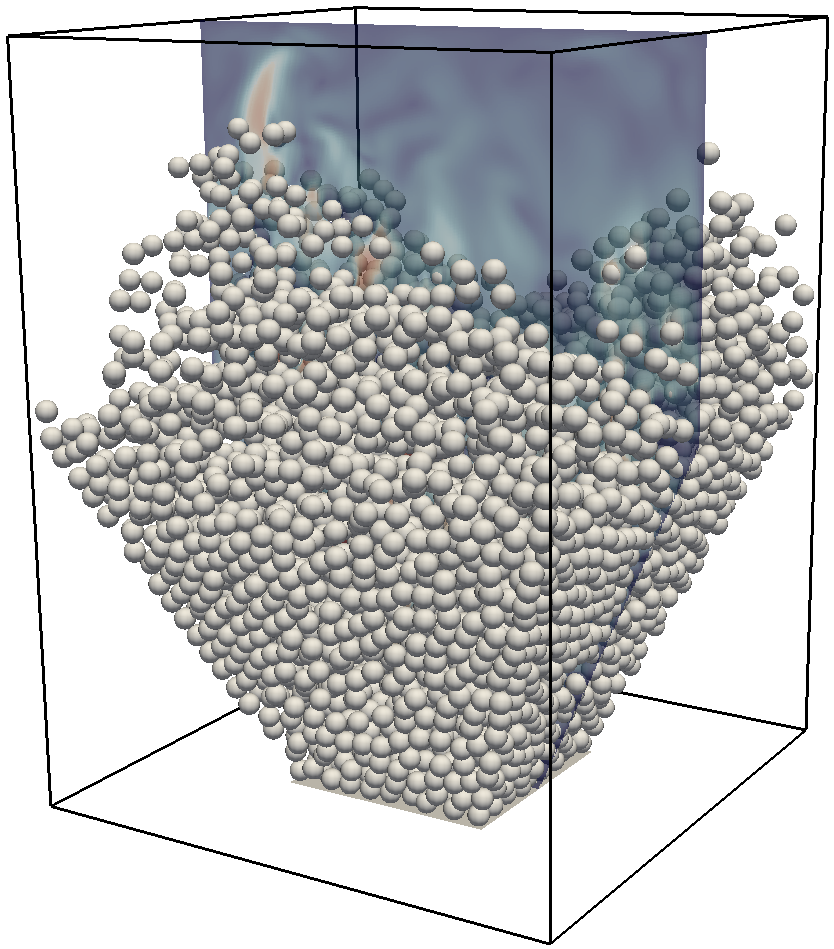}
		\caption{Active settling phase.}
	\end{subfigure}
	~
	\begin{subfigure}[t]{0.32\textwidth}
		\centering
		\includegraphics[width=\textwidth]{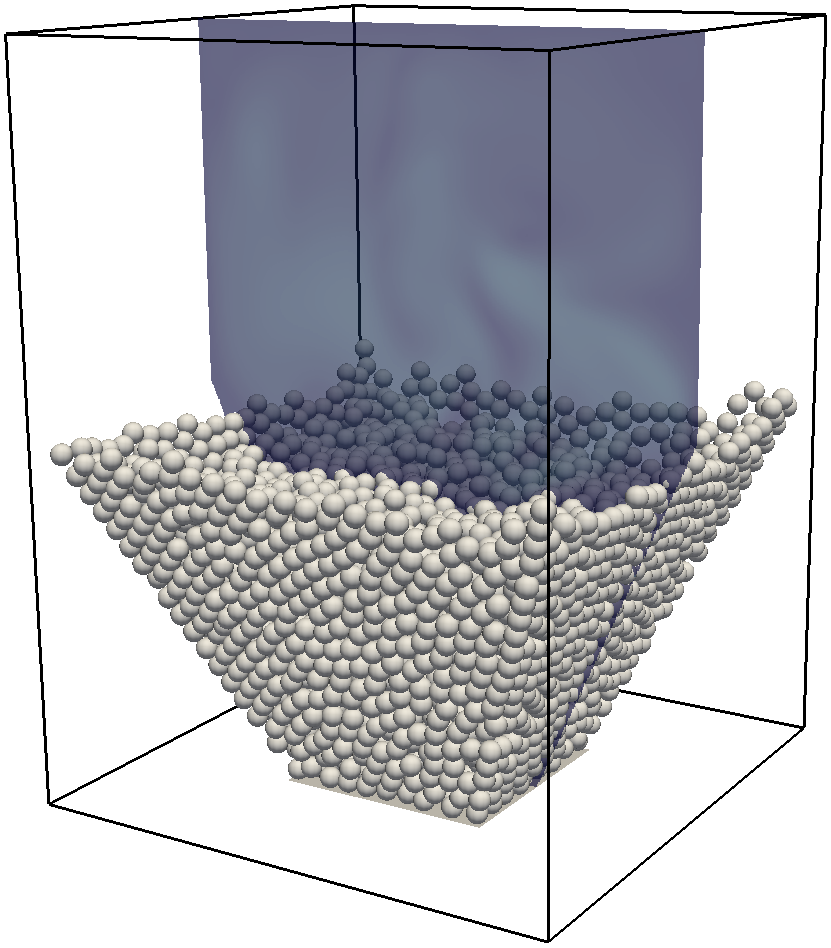}
		\caption{Final, completely settled state.}
	\end{subfigure}
	\caption{Visualization of the temporal evolution of the particles and the fluid velocity in the hopper clogging simulation.}
	\label{fig:hopper_sim}
\end{figure}

The simulation domain is a rectangular box that is subdivided into $12\times12\times16$ blocks of size $32^3$ cells.
This results in a total of 2304 blocks.
By using $N_p=256$ processes of the SuperMUC cluster, a uniform block distribution is achieved with 9 blocks per process.
We use four inclined planes to construct a fluid domain where the horizontal cross section is reduced by 60\% towards the bottom plane, resembling a hopper as illustrated in Figure~\ref{fig:hopper_sim}.
As a result, several blocks remain empty throughout the whole simulation.
The domain contains about $4300$ spherical particles of diameter $D/\Delta x = 15$ that are initially densely clustered along the top plane and then settle under the effect of gravity.
The Galileo number is $Ga=50$, the density ratio is $\rho_p/\rho_f=1.5$, the number of rigid body solver sub cycles is set to $S=10$, and the simulation is run for $80000$ time steps.
The temporal evolution of the simulation can be seen in Figure~\ref{fig:hopper_sim}.

In order to distribute the blocks among the available processes based on the predicted loads, we use several of the strategies presented in Sec.~\ref{sec:load_dist_strategies} and evaluate their performance.
For space-filling curves, Hilbert and Morton orders are applied and the mentioned ParMETIS variants are also included.
In all cases, the load balancing step that consists of load estimation and load distribution is triggered every $100$ time steps.
These variants are compared against the case where no load balancing is applied at all.
However, since the initial distribution of $9$ blocks per process is non-trivial, we use the static partitioning given by the Hilbert curve for that case, which is thus called once at the beginning of the simulation.

\subsection{Results of the Hopper Simulation}

\begin{figure}[t]
	\centering	
	\input{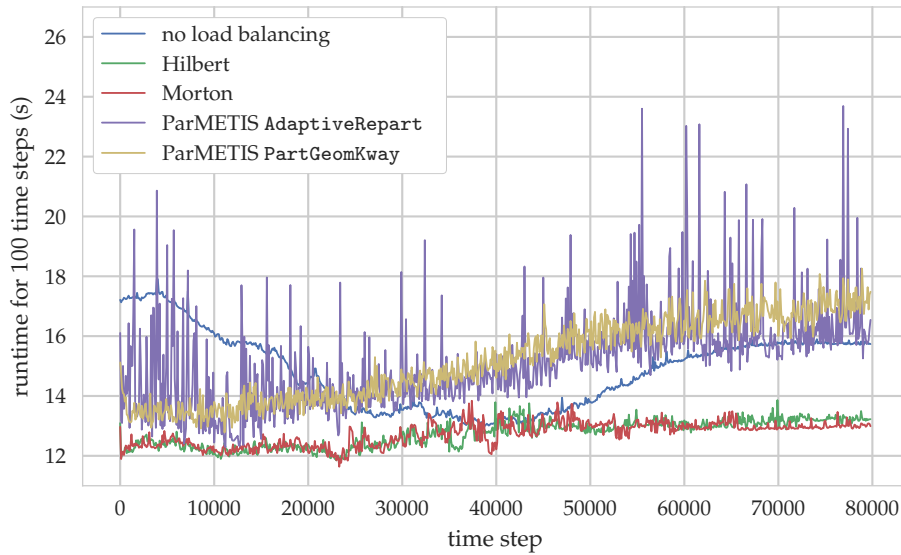}
	\caption{Runtime per 100 time step in seconds over the time steps for simulation shown in Figure~\ref{fig:hopper_sim}. Four load distribution variants using the load estimator from Sec.~\ref{sec:load_determination} are shown together with the case where no load balancing is applied.}
	\label{fig:sim_rtpts}
\end{figure}

\begin{table}[t]
	\caption{Comparison of total runtime of the hopper simulation for the four load distribution variants relative to the case without load balancing. Additionally, the runtime of the load balancing step relative to the total runtime and the simulation-averaged edge cut is given.}
	\label{tab:sim_rt_comp}
	\centering
	\begin{tabular}{l|c|c|c}
		\toprule
		variant & relative runtime (\%) & load balancing fraction (\%) & mean edge cut\\
		\midrule
		no load balancing & 100.0 & 0.0 & $4.5\cdot10^6$\\
		Hilbert & 86.0& 2.0 & $4.3\cdot10^6$\\
		Morton & 86.0 & 1.9 & $4.8\cdot10^6$\\
		ParMETIS \texttt{AdaptiveRepart} & 101.9 & 3.9 & $3.6\cdot10^6$\\ 
		ParMETIS \texttt{PartGeomKway} & 102.3 & 6.8 & $3.6\cdot10^6$\\
		\bottomrule
	\end{tabular}
\end{table}

In Figure~\ref{fig:sim_rtpts}, the temporal evolution of the runtime for $100$ time steps is shown for the four different load distribution variants and compared against the case with no load balancing.
From these results, we see that without load balancing the runtime decreases until time step $40000$, which is shown in the middle of Figure~\ref{fig:hopper_sim}.
This corresponds to the transition from an initially densely packed bed to a fully suspended settling with only few inter-particle collisions. 
Afterwards, when the clustering at the bottom sets in, the simulation again gets slower. 
On the other hand, both cases that use space-filling curves for load distribution yield a relatively constant runtime throughout the whole simulation and are both very similar in their behavior.
Except for some fluctuations halfway through the simulation, the space-filling curves are significantly faster than the default case.
The two ParMETIS variants also feature low runtimes at the beginning which then steadily increase as the simulation advances, leading to larger runtimes than the default case.
Again, the trend of both curves is rather similar but the \texttt{AdaptiveRepart} variant exhibits stronger oscillations.

The time to solution of the whole simulation is evaluated as the cumulated sum of the values depicted in Figure~\ref{fig:sim_rtpts} and is summarized in Table~\ref{tab:sim_rt_comp}.
In total, load balancing with space-filling curves reduces the overall runtime by 14\% whereas using ParMETIS even results in a slight increase of the runtime.
Compared to the rather simple space-filling curves, the more complex load distribution algorithms of ParMETIS are also reflected in the larger fraction of the runtime spent in the load balancing procedure.
As a result, however, the ParMETIS variants show the smallest edge cut, i.e. the sum of all edge weights from Eq.~\eqref{eq:edge_weights} that are cut by process boundaries.

\begin{figure}[t]
	\centering	
	\input{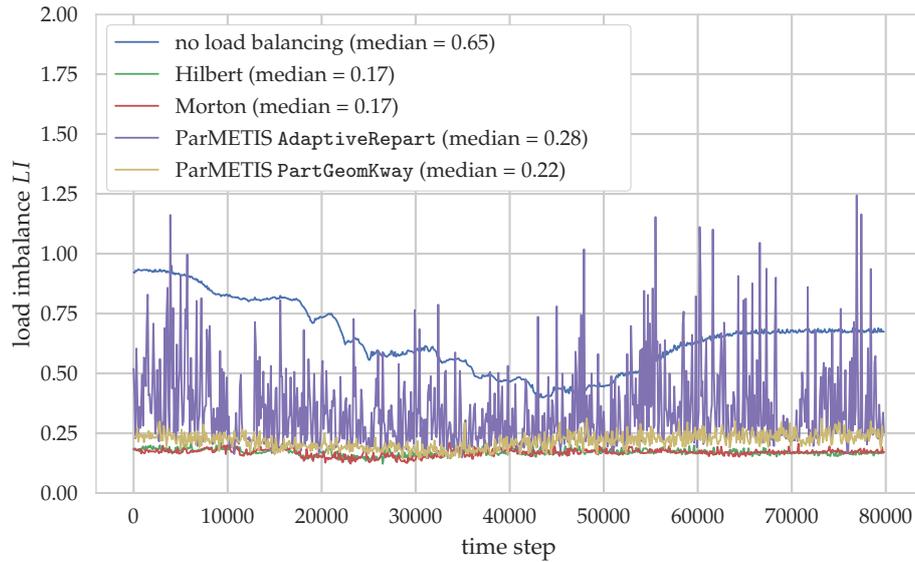}
	\caption{Load imbalance over the time steps for simulation shown in Figure~\ref{fig:hopper_sim}.}
	\label{fig:sim_loadimbal}
\end{figure}

Since the overall goal of load balancing is to reduce possible load imbalances, the latter ones are again evaluated in intervals of $100$ time steps and shown in Figure~\ref{fig:sim_loadimbal}.
The load imbalance $LI$, ideally close to zero, is here defined as the normalized difference between the maximum and the average runtime per interval, i.e. 
\begin{equation}
LI = \frac{\max_p m_{tot}(p)}{\tfrac{1}{N_p}\sum_p m_{tot}(p)} - 1,
\end{equation}
where $m_{tot}$ is the runtime of the workload generating parts, see Eq.~\eqref{eq:mtot}, summed up over the $100$ time steps, and $p$ denotes one of the $N_p=256$ processes.
This definition takes into account that the performance of the simulation is most affected by a few processes that are slower than average, resulting in many idle processes, whereas having some processes that are faster than average is less problematic, as only a few processes have to wait.
Our definition thus differs from \cite{owen2000} who used the normalized difference between the maximum and minimum runtime. 
The evolution of the load imbalance for the case without load balancing resembles the behavior from Figure~\ref{fig:sim_rtpts} with a drop from 90\% to around 40\%, followed by an increase to 70\%.
The space-filling curves exhibit the lowest imbalances with a median of 17\%, and thus almost four times smaller than the default case.
Two distinct behaviors can be seen for ParMETIS: the \texttt{PartGeomKway} variant is slightly worse than the space-filling curves with an median of 22\% whereas the \texttt{AdaptiveRepart} case shows larger imbalances together with strong fluctuations. 

\subsection{Discussion}

The results from this comparative study show the importance of choosing an appropriate load distribution technique.
Space-filling curves achieve a reduction of the overall runtime by 14\% compared to a simulation in which no load balancing is applied.
On the other hand, ParMETIS, which relies on space-filling curves but combines them with more advanced partitioning strategies, does not yield an improvement in the runtime.
Even though the ParMETIS variants show a reduction of the load imbalance and the edge cut, they ultimately fail in reducing the time to solution.
One reason for that is certainly but only partly that the runtimes for the load balancing step are significantly higher than for the space-filling curves.
Also, some improvements can be expected by optimizing the parameters that have to be specified for the ParMETIS routines.
This specialization, however, runs contrary to a widely and readily applicable load distribution technique.
Another reason is that ParMETIS, like many other graph partitioning tools, is not designed to work with an average of, here, $9$ blocks per process but usually expects around $1000$ or more.
This would e.g. be the case for cell-based domain partitioning, but unrealistic for block-based ones, and with this larger granularity the partitioning routines are expected to work faster and better.
The observable strong fluctuations in the runtime and load imbalance graphs can possibly addressed to be a direct cause of this difficulty.
Lastly, our choice of the edge weights, Eq.~\eqref{eq:edge_weights}, can be improved as it currently is solely based on the volume of the communication of the LBM data structures.
Thus, the communication by the particle simulation is not regarded but might change the weights notably.
One could, therefore, also try to find an estimator for the edge weights similarly to the procedure used to come up with a vertex weight estimator.

We note that these results are specific for the here simulated setup and cannot be generalized easily.
They also show, however, that load balancing is a viable tool to reduce the time to solution if carefully applied and systematically compared.
It is especially helpful for cases where strong heterogeneities exist in the simulation, as it is the case for the initial and end phase of the here carried out simulation.

\section{Conclusion}
\label{sec:conclusion}

In this work, we present and evaluate different techniques to improve the performance of particulate flow simulations that exhibit spatially and temporally varying workloads during their parallel execution.

To this end, first a workload estimator is designed and calibrated that predicts the workload for each block of the domain using locally available information.
It is based on analyzing the involved algorithms and setting up suitable functions that describe the generated load.
The coefficients of the functions are determined by fitting them to timing measurements obtained from simulations of a small but representative setup.
Choosing such a setup is a crucial step as all relevant phenomena and properties that can lead to varying workloads must be included to obtain a generally applicable estimator.
Since timing measurements are influenced by various factors, like the hardware used or the actual implementation details, major changes in these might require a reevaluation of the fitted coefficients.
Even though the article focuses on particulate flow simulations, the presented approach can be readily used for other multiphysics simulations.

In a second step, this workload estimator is applied in a more complex simulation scenario in conjunction with a load distribution technique to reduce load imbalances between the processes.
We compare and evaluate load distribution via space-filling curves and with the help of the graph partitioning tool ParMETIS to the case without any load balancing.
For the case of space-filling curves, a significant reduction of the time to solution by 14\% can be observed, demonstrating how these techniques can be used to utilize hardware resources more efficiently by minimizing load imbalances.
On the other hand, the ParMETIS-based distributions are unable to improve the runtime for this specific scenario.
This can be attributed to the relatively large computational overhead produced by the load balancing step but also generally to the different focus of these graph partitioning tools that would require a finer granularity. 

All in all, this study shows that load balancing is an effective tool for parallel multiphysics simulations where workloads change in time and space.
It is typically used in combination with adaptive mesh refinement, where it becomes a necessity due to the strongly varying work loads when the grid is updated.
Thus, naturally, we aim to apply the here presented load balancing approach with AMR.
But, as we have shown, it can also be advantageous to apply it with uniform grids to reduce the time to solution of the simulation. 

Future work will pursue further improvements of the load estimator as well as the load distribution step. 
By including information available from performance models for the different algorithms, the workload estimator can be made more general and flexible. 
Tools like Kerncraft~\cite{Kerncraft} automatically analyze the performance of a given implementation for the hardware at hand, which would render the estimator independent of these factors.
Furthermore, a workload estimate based on the current runtimes is a natural alternative to the proposed predictor as it is able to use actual data from the currently running simulation.
Such an estimator can not be used in an a-priori fashion which renders it less useful for adaptive grid refinement where new blocks are created and have to be distributed immediately to avoid huge imbalances.
However, a combination of both strategies would result in a kind of predictor-corrector approach and could reduce load imbalances more effectively. 
For the load distribution step, other graph partitioning libraries are available, e.g. Zoltan~\cite{ZoltanIsorropiaOverview2012}, PT-Scotch~\cite{CHEVALIER2008318} or Geographer~\cite{von2018balanced}.
In combination with a better estimator for the communication weights, their performance should thus be evaluated and again be compared to the space-filling curves.


\vspace{6pt} 


\authorcontributions{ Christoph Rettinger developed the presented algorithms, implemented them, realized the simulations and evaluated the results. Furthermore, he wrote major parts of this article. Ulrich Rüde supervised the work and contributed to the arrangement of this article by significantly improving the structure and wording of this manuscript. }

\funding{ This research received no external funding.}

\acknowledgments{The first author would like to thank Sebastian Eibl for valuable input and discussions. Both authors gratefully acknowledge the Gauss Centre for Supercomputing e.V. (\url{www.gauss-centre.eu}) for funding this project by providing computing time on the GCS Supercomputer SuperMUC at Leibniz Supercomputing Centre (\url{www.lrz.de}). They also acknowledge support by Deutsche Forschungsgemeinschaft and Friedrich-Alexander-Universität Erlangen-Nürnberg (FAU) within the funding programme Open Access Publishing.}

\conflictsofinterest{ The authors declare no conflict of interest.} 

\externalbibliography{yes}
\bibliography{Library}

\end{document}